\documentclass[twocolumn]{aastex6}

\newcommand\Jup{\ensuremath{_{\mathrm{\scriptscriptstyle Jup}}}} 
 
\newcommand\EUV{\ensuremath{_{\mathrm{\scriptscriptstyle EUV}}}}

\begin{document}


\title{ALMA Observations of the Extraordinary Carina Pillars: HH 901/902}




\author{Geovanni Cortes-Rangel, Luis A. Zapata, Jes\'us A. Toal\'a}
\affiliation{Instituto de Radioastronom\'\i a y Astrof\'\i sica, Universidad Nacional Aut\'onoma de M\'exico, \\P.O. Box 3-72, 58090, Morelia, Michoac\'an, M\'exico.}


\author{Paul T. P. Ho}
\affiliation{Academia Sinica Institute of Astronomy and Astrophysics, PO Box 23-141, Taipei, 10617, Taiwan. \\ East Asian Observatory, 666 N. A'ohoku Place, Hilo, Hawaii 96720, USA.}

\author{Satoko Takahashi}
\affiliation{Joint ALMA Observatory, Alonso de Cordova 3108, Vitacura, Santiago, Chile. \\ NAOJ Chile Observatory, Alonso de Cordova 3108, Vitacura, Santiago, Chile. \\
Department of Astronomical Science, School of Physical Sciences, SOKENDAI, Mitaka, Tokyo 181-8588, Japan.}

\author{Adal Mesa-Delgado}
\affiliation{Instituto de Astrof\'\i sica, Facultad de F\'\i sica, Pontificia Universidad Cat\'olica de Chile, \\Av. Vicu\~{n}a Mackenna 4860, 782-0436 Macul, Santiago, Chile. }

\author{Josep M. Masqu\'e}
\affiliation{Departamento de Astronom\'\i a, Universidad de Guanajuato, Apdo. Postal 144, 36000 Guanajuato, M\'exico.}




\begin{abstract}
We present Atacama Large Millimeter/Submillimeter Array (ALMA) 1.3 mm continuum 
and C$^{18}$O(2$-$1), N$_2$D$^{+}$(3$-$2), $^{13}$CS(5$-$4), and $^{12}$CO(2$-$1) line sensitive 
and high angular resolution ($\sim$0.3$''$) observations of the famous carina pillars 
and protostellar objects HH 901/902. Our observations reveal for the first time, the bipolar CO 
outflows and the dusty disks (plus envelopes) that are energizing the extended and irradiated HH 
objects far from the pillars. We find that the masses of the disks$+$envelopes are about 
0.1 M$_\odot$ and of the bipolar outflows are between 10$^{-3}$ - 10$^{-4}$ M$_\odot$, which suggests that
they could be low- or maybe intermediate- mass protostars. 
Moreover, we suggest that these young low-mass stars are likely embedded 
Class 0/I protostars with high-accretion rates. We also show the kinematics of the gas in the pillars together 
with their respective gas masses (0.1 -- 0.2 M$_\odot$).  We estimate that the pillars will be photo-evaporated 
in 10$^4$ to 10$^5$ years by the massive and luminous stars located in the Trumpler 14 cluster.  
Finally, given the short photo-evaporated timescales and that the protostars in these pillars are still very embedded, 
we suggest that the disks inside of the pillars will be quickly affected by the radiation of the massive stars, forming proplyds,
like those observed in Orion. 
\end{abstract}




\section{Introduction} \label{sec:intro}

The Carina Nebula is one most active star forming regions in our Galaxy. This region 
contains nearly one hundred O-type stars and tens of thousands of lower-mass young 
stars \citep{wal1973,wal1995,mas1993,smi2006,pov2011,pre2011,fei2011}, with the Trumpler (Tr) 14 and 16 
massive clusters being one the most 
luminous regions within Carina. The clusters Tr 14 and 16 have ultraviolet (UV) luminosities 
about 20 and 60 times higher than $\theta$1 Ori C located in the Orion Nebula \citep{sb2008}.
Hence, the Carina Nebula is an ideal laboratory to study the time-scale of the photo-evaporation of
disks, envelopes, and molecular pillars located especially in the core of the massive clusters.  

\begin{figure*}[!]
\begin{center}
\includegraphics[scale=0.74]{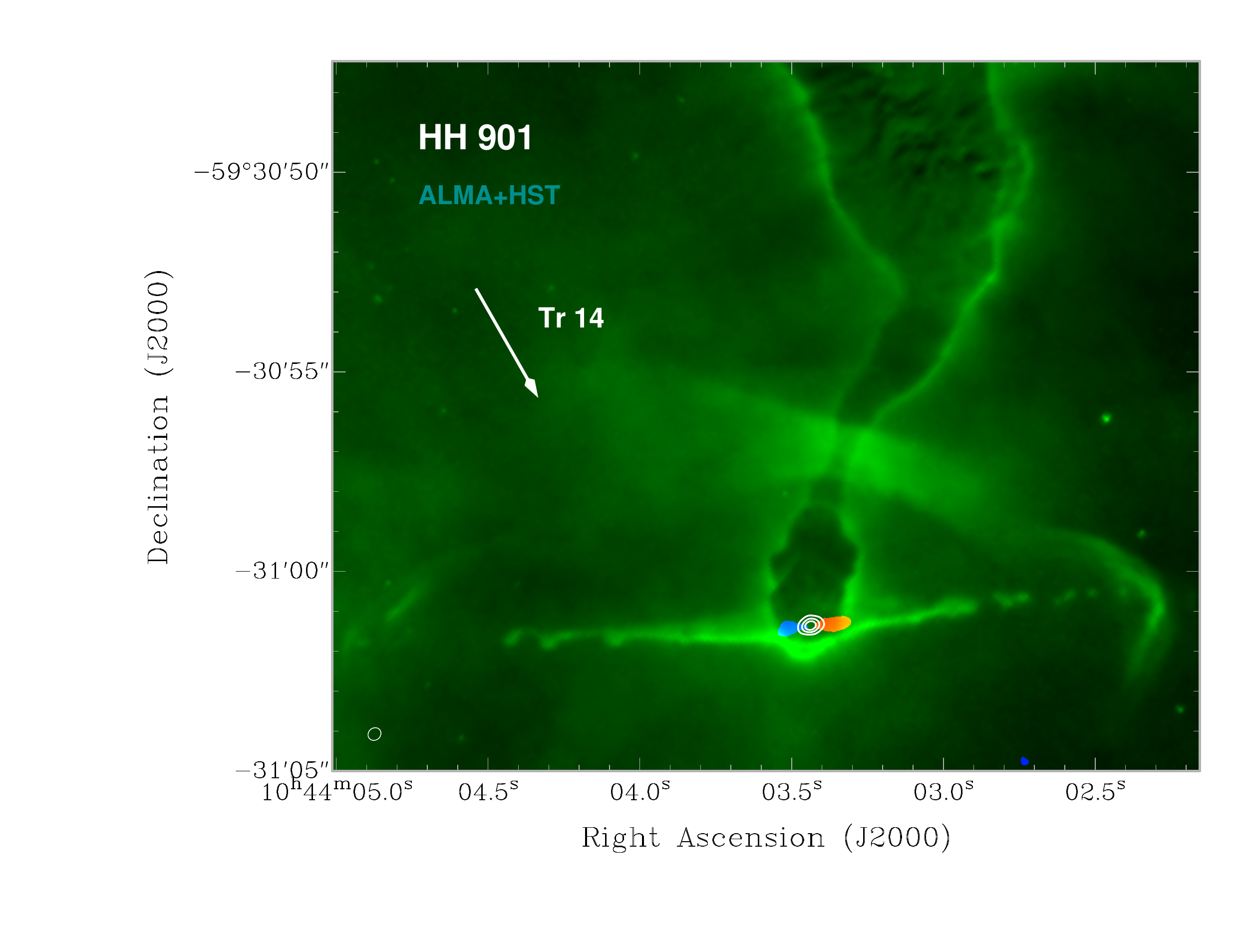}
\caption{ALMA CO(2$-$1) moment zero (blue and red colors) and millimeter continuum (contours) images 
of the HH 901 object overlaid in a {\it HST}(H$\alpha$) optical image (green colors). The blue and
red color represent the blue-shifted and red-shifted CO emission, respectively, from
the HH 901 outflow. The contours range from 50\% to 90\% of the peak emission, 
in steps of 20\%. The peak of the millimeter continuum emission is 2.8 mJy Beam$^{-1}$.
Here, we are only contouring the most compact 1.3 mm emission from our observations revealing 
the envelope and the disk. There is still some extended emission at low levels.   
The half-power contour of the synthesized beam of the line image is shown in the bottom-left corner.
We integrate in radial velocities from $+$6.4 to $-$1.2 km s$^{-1}$ for the redshifted emission, 
and from $-$8.2 to $-$14.6 km s$^{-1}$ for the blueshifted emission for the CO outflow in the HH 901 object. 
The LSR systemic velocity of the entire pillar associated with the object HH 901 is about $-$5.0 km s$^{-1}$.}
\label{f1}
\end{center}
\end{figure*}

As a clear case of irradiated/photo-evaporating objects, using APEX FLASH$+$ and  
CHAMP$+$ line observations, \citet{sah2012} reported  an irradiated object with a cometary-shaped 
(CN 105$-$600) in the Carina star-forming region. They found
 that this externally irradiated object is not a classical proplyd as those reported in the Orion Nebula 
 \citep{ode1993, ode2008, hen1999}, this object is instead a large and massive molecular globule 
 (0.1 -- 0.3 M$_\odot$) in which very compact disks and outflows are embebed. The young star inside of this globule has
 a mass of about 0.5 M$_\odot$  and total luminosity of about 10 L$_\odot$ \citep{sah2012},
 corresponding to a low-mass young star.  \citet{hai2017} also reported APEX observations of several
 globulettes in Carina, these globulettes are compact and denser than objects of similar 
 mass in the Rosette nebula.   
 
            There are however many more well studied irradiate globules in the Carina Nebula, they include to HH 666, 
            HH 1066, HH 900, and HH 1010 \citep{rei2013}. But any of them include interferometric millimeter observations. 
            Given that different young massive star forming regions have a very varied morphology, ages, and number of massive stars, 
            these regions hold distinct kinds of irradiated interfaces (or globulettes) and their study is important because this will allow 
            us to better understand the evolution of the young stars. 
            A clear example is found in the Carina region \citep{har2012}, which contains a rich array of walls, pillar and globules
            and in the older Cyg OB2 association is found mostly isolated globules. 

Very recently,  \citet{del2016} using ALMA (Atacama Large Millimeter/Submillimeter Array) Band 6 Long Baseline continuum observations 
reported the first direct imaging of protoplanetary disks in the star-forming region of Carina and studied 
their possible external strong photo-evaporation. They detected a few protoplanetary disks (CN104-593 and CN105-600)
in the highly obscured globules within the Carina Nebula and found no signs of disks 
near to the massive clusters Tr~14 and Tr~16. 
The disks have an average size of 120~au and total masses of 30 and 50~\(M\Jup\).
The non-detection of millimeter emission above the 4$\sigma$ threshold (\(\sim 7 M\Jup\)) in the core of the massive cluster Tr~14, 
suggest evidence for rapid photo-evaporative disk destruction in the cluster's harsh radiation field.  

           Bipolar molecular outflows and collimated optical jets are one of the phenomena associated with star formation \citep{ball2016}.
           The outflows are composed of swept-up molecular shells energized by neutral and atomic jets containing molecules. As these molecular outflows 
           escape from their parent molecular clouds to scales of some parsecs, they become mostly atomic or ionized and traced by Herbig-Haro (HH) objects
           revealed at optical wavelengths. 

The HH 901/902 were first reported by \citet{smi2010} using {\it Hubble Space Telescope (HST)}/Advanced 
Camera for Surveys  H$\alpha$ observations. They catalogued the HH objects as irradiated bipolar jets. These objects 
are extremely bright and clear in their {\it HST} images (see their Figure \ref{f5}). The dusty pillars from which HH 901 and 902 emerge
are between Tr 14  and 16 clusters.  Both HH objects axis are clearly perpendicular to the dusty pillars and are 
located to a distance of about 1.5 arcmin. to the south of the Tr~14 and about 7 arcmin. from Tr~16.  These flows consist of  
a highly collimated (with opening angles of only a few degrees) chain of H$\alpha$-emitting knots 
with an extent of 16 arcsec (0.18 pc) for the HH 901 and an extent of 38 arcsec (0.42 pc) for the HH 902. 
Both flows are strongly irradiated by the massive clusters Tr 14 and 16.
HH 901 and HH 902 both lie close to plane of the sky with tilt angles of about 20$^\circ$ \citep{rei2014}.
\citet{rei2013,rei2014} suggested that these two jets are energized by intermediate-mass stars (2-8 M$_\odot$).
However, these studies could not corroborate this interpretation due to the fact that the 
authors reported that the driving sources were not detected in the IR, probably because they were too 
embedded and because of inadequate angular resolution. 
Given these problems, the fits to the IR SEDs (Spectral Energy Distributions) become very poor and the estimated luminosity is uncertain. 

The HH 901/HH 902 objects have also been detected in H$_2$ (2.2 $\mu m$) and Br$\gamma$ spectral lines by \citet{har2015}.
These observations represented the first detection of the molecular material associated with the HH objects. In particular, the H$_2$ emission 
is tracing the innermost parts of the outflow as can be seen in Figures 19 and 20 of  \citet{har2015}. Molecular material far from the 
exciting objects is probably destroyed by the UV radiation.

In this study, we have carried out new ALMA observations of the HH 901/HH 902 objects in order to better characterize 
their optical jets, disks, and molecular pillars. We reveal their compact molecular outflows, dusty
disk, and envelopes, together with the kinematics of the pillars. We describe these results in more detail in the next sections.

\begin{figure}
\centering
\includegraphics[scale=0.54]{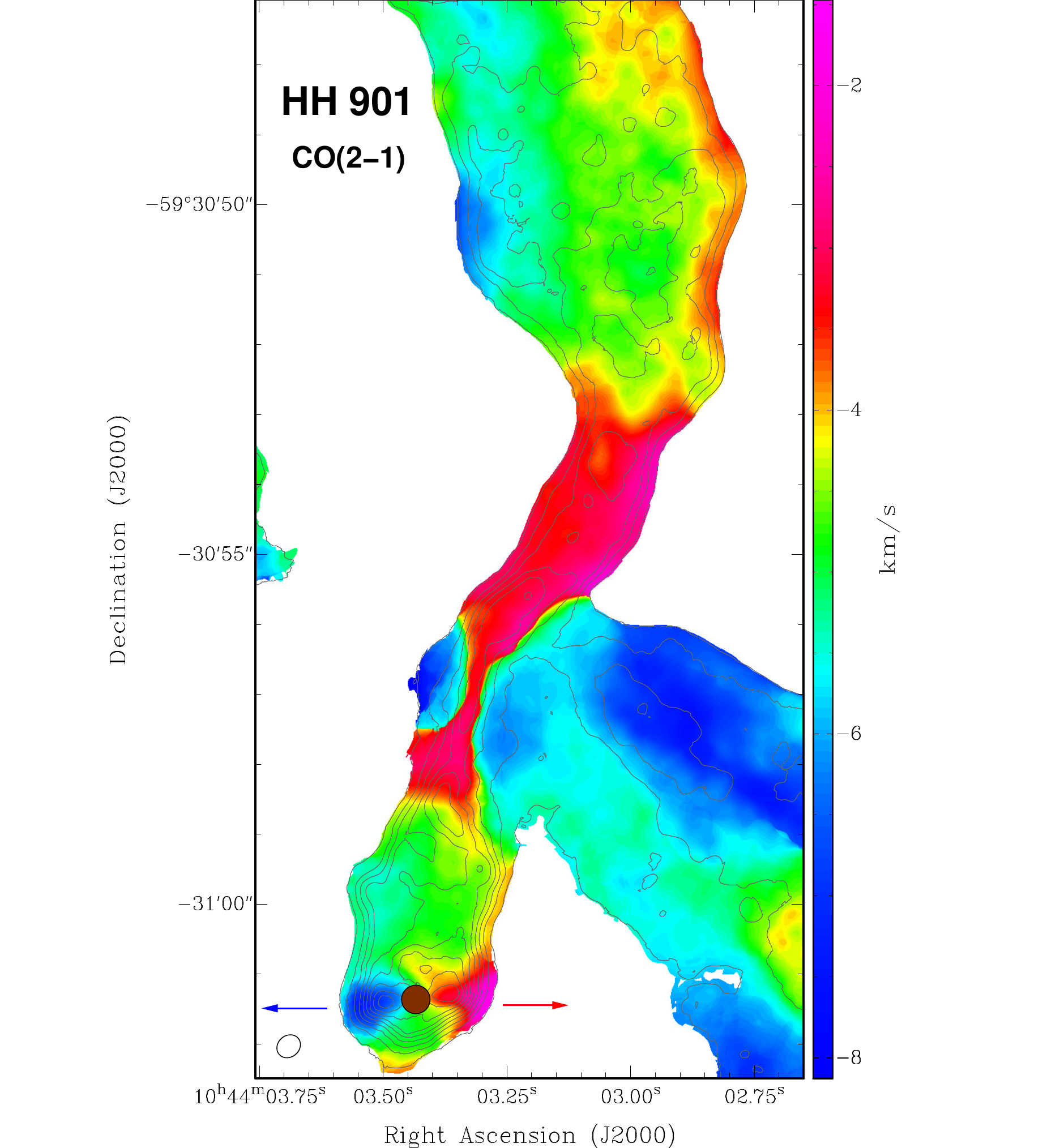}
\caption{ALMA CO(2$-$1) moment zero (contours) and one (colors) maps of the pillar and outflow HH 901. 
The contours range from 10\% to 90\% of the peak emission, in steps of 10\%. The peak of the millimeter 
line CO emission is 1.8 Jy Beam$^{-1}$ km s$^{-1}$. The half-power contour of the synthesized beam of the line image 
is shown in the bottom-left corner. The LSR radial velocity scale-bar is shown at the right. The brown circle
marks the position of the disk revealed in these ALMA observations for the HH 901 object.
The red and blue arrows trace the orientation of the outflow. For the CO moment zero image,
we integrate in radial velocities from $-$15.9 to $+$6.4 km s$^{-1}$. 
The LSR systemic velocity of the entire pillar associated with the object HH 901 is about $-$5.0 km s$^{-1}$.}
\label{f2}
\end{figure}

\begin{figure*}
\centering
\includegraphics[scale=0.63]{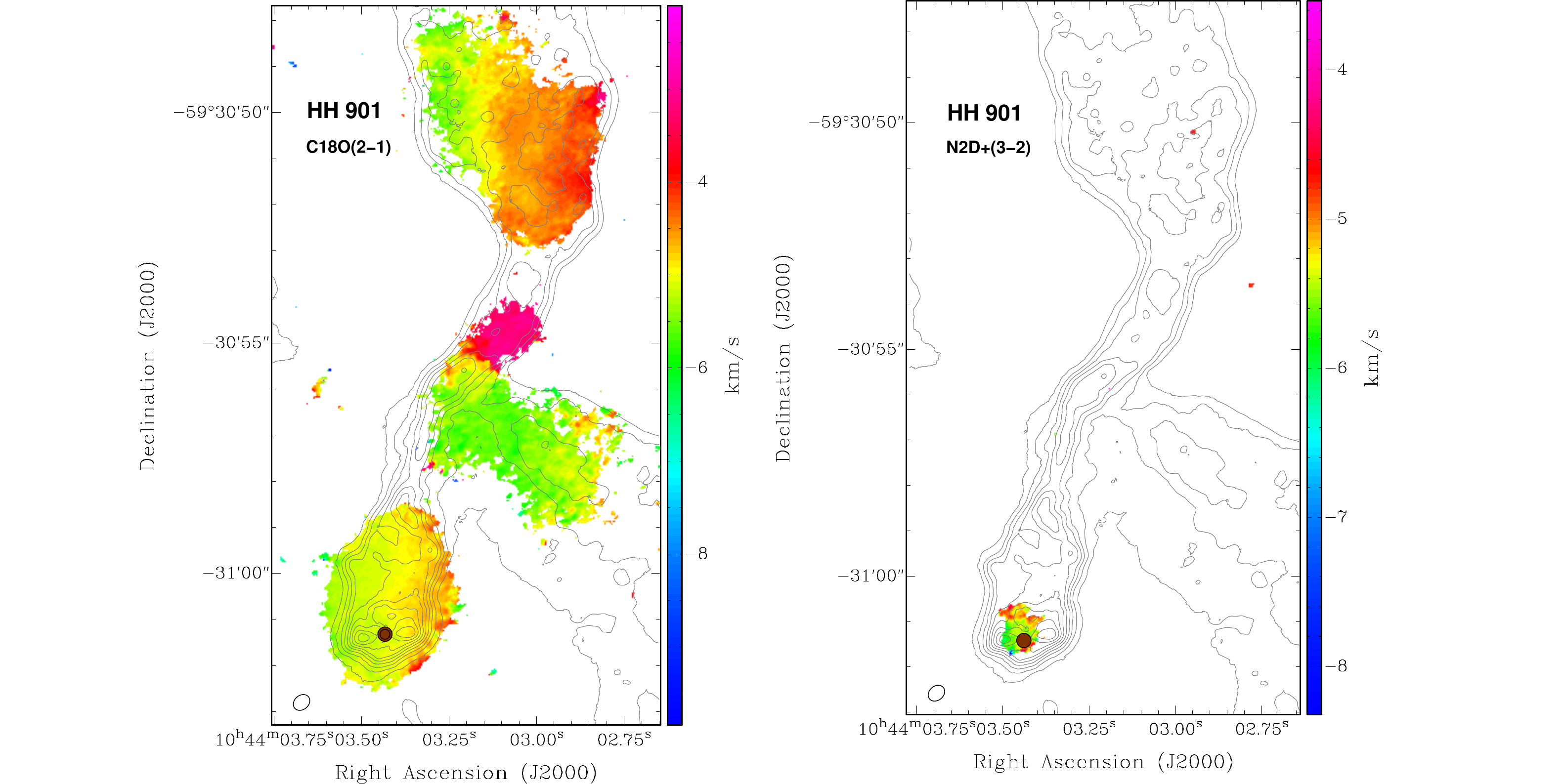}
\caption{ALMA C$^{18}$O(2$-$1) (left panel) and N$_2$D$^{+}$(3$-$2) (right panel) moment one maps overlaid 
with a contour image of the CO(2$-$1) moment zero of the pillar and outflow HH 901. 
The contours range from 10\% to 90\% of the peak emission, in steps of 10\%. The peak of the millimeter 
line CO emission is 1.8 Jy Beam$^{-1}$ km s$^{-1}$. The half-power contour of the synthesized beam of the line image 
is shown in the bottom-left corner of each panel. The LSR radial velocity scale-bar is shown at the right. The brown circle
marks the position of the disk$+$envelope revealed in these ALMA observations for the HH 901 object.
}
\label{f3}
\end{figure*}

\section{Observations} \label{sec:obs}

The observations of the HH 901/902 protostellar objects were carried out with
the ALMA at Band 6 in 2018 January 1st and 4th (C43-6), and April 20th (C43-3) 
as part of the Cycle 5 program 2017.1.00912.S.
The observations used 43 antennas with a diameter of 12 m, yielding baselines 
with projected lengths from 15 $-$ 2516 m (11.5 $-$ 1935 k$\lambda$) for C43-6 and
15 $-$ 500 m (11.5 $-$ 384.6 k$\lambda$) for C43-3. The primary beam at this frequency 
has a full width at half-maximum (FWHM) of about 25$''$,  so that each object was covered 
 with a single pointing at the sky position  $\alpha(J2000) = 10^h~ 44^m~ 1\rlap.^s7$;
$\delta(J2000) = -59^\circ~ 30'~ 32.0''$ for the HH 902, and  
$\alpha(J2000) = 10^h~ 44^m~ 3\rlap.^s5$; $\delta(J2000) = -59^\circ~ 31~ 2.0''$ for the HH 901.
The integration time on source (HH 901 and 902) for the C43-6 configuration was about 9 min. and 
for the C43-3 configuration is about 3.5 min. 

\begin{figure*}
\begin{center}
\includegraphics[scale=0.55]{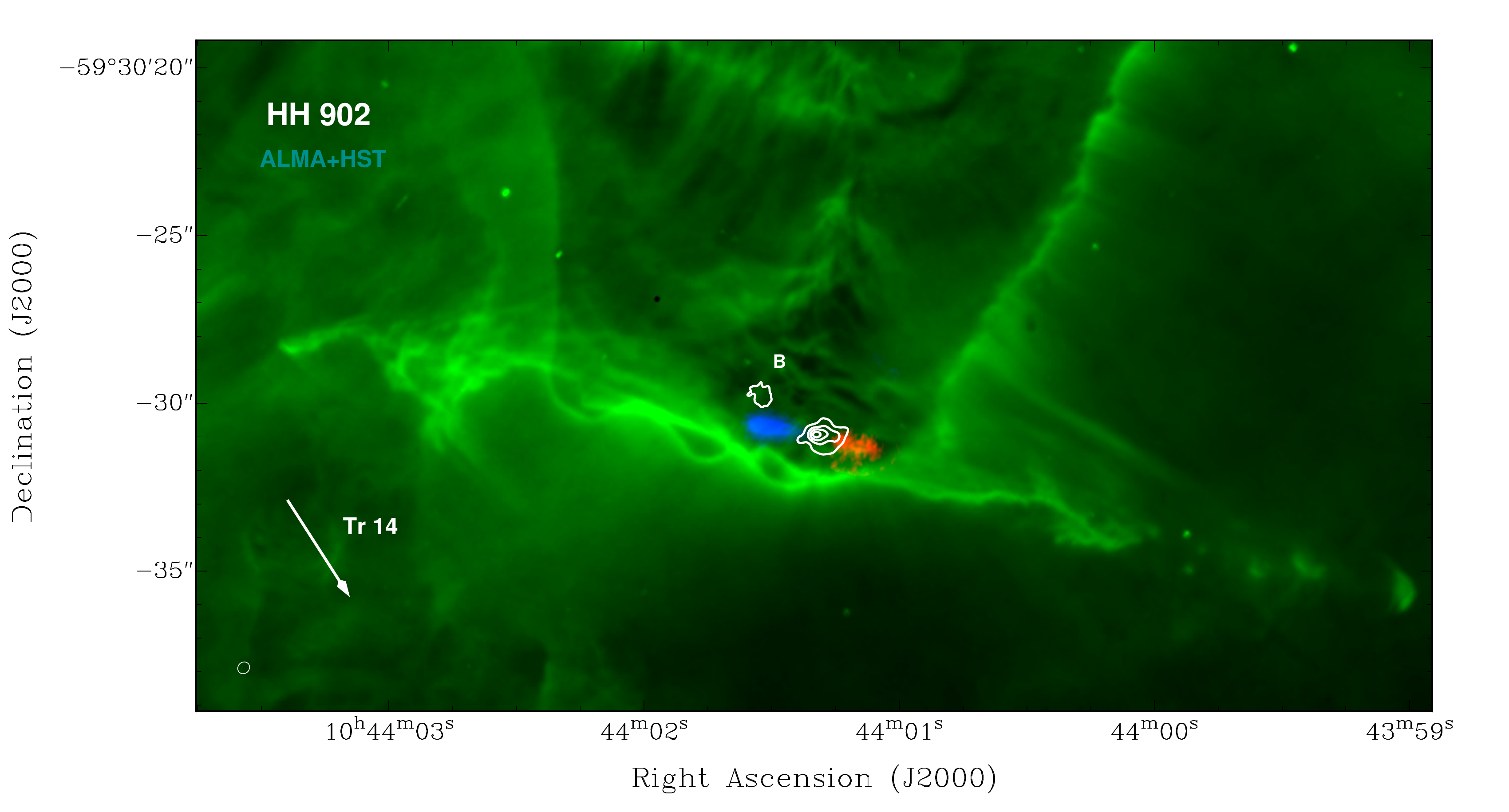}
\caption{ALMA CO(2$-$1) moment zero (blue and red colors) and millimeter continuum (contours) images 
of the HH 902 object overlaid in a {\it HST} optical image (green colors). The blue and
red color represent the blue-shifted and red-shifted emission, respectively, from
the HH 902 outflow. The contours range from 50\% to 80\% of the peak emission, 
in steps of 10\%. The peak of the millimeter continuum emission is 5.1 mJy Beam$^{-1}$.
Here, we are only contouring the most compact 1.3 mm emission from our observations revealing 
the envelope and the disk. There is still some extended emission at low levels.   
The half-power contour of the synthesized beam of the continuum image is shown in the bottom-left corner.
We integrate in radial velocities from $-$3.7 to $-$5.6 km s$^{-1}$ for the redshifted emission, 
and from $-$10.8 to $-$15.2 km s$^{-1}$ for the blueshifted emission for the CO outflow. 
The LSR systemic velocity of the entire pillar associated with the object HH 902 is about $-$8.5 km s$^{-1}$.}
\label{f4}
\end{center}
\end{figure*}

The continuum images were obtained by averaging line-free spectral channels of four
spectral windows centered at:  218.014 GHz(spw0), 219.578 GHz(spw1), 231.239 GHz(spw2),
and 230.556 GHz(spw3).  The total bandwidth for the continuum is about 4.3 GHz.  
These four spectral windows were centered to observe different molecular lines 
as the C$^{18}$O(2$-$1) ($\nu_\mathrm{rest}$= 219.56035 GHz), $^{13}$CS(5$-$4) ($\nu_\mathrm{rest}$=231.22068 GHz), 
N$_2$D$^{+}$(3$-$2) ($\nu_\mathrm{rest}$=231.32186 GHz), and $^{12}$CO(2$-$1) ($\nu_\mathrm{rest}$=230.53800 GHz).
We detected emission only from the lines C$^{18}$O,  N$_2$D$^{+}$, and
$^{12}$CO with a channel spacing of 0.63 km s$^{-1}$.  The thermal emission
from these species can be seen in Figures \ref{f1}, \ref{f2}, \ref{f3}, \ref{f4}, \ref{f5}, and \ref{f6}.

The weather conditions were very good for these observations and stable with an average 
precipitable water vapor between 1.3 and 2.2 mm and an average system temperature 
around 90 K. The ALMA calibration included simultaneous observations of the 183 GHz 
water line with water vapor radiometers, used to reduce atmospheric phase fluctuations.
Quasars, J0904-5735 and J1107-4449 were used for the bandpass and flux calibrations. 
J1032-5917 was used for correcting the gain fluctuations. 

The data were calibrated, imaged, and analyzed using the Common Astronomy Software Applications (CASA)
Version 5.1. Imaging of the calibrated visibilities was done using the CLEAN and TCLEAN tasks. 
We concatenated the data from both dates for the C43-6 and C43-3 configurations with the CONCAT task.
We used the ROBUST parameter of CLEAN in CASA set to natural.
 The resulting image rms noise for the continuum was 50 $\mu$Jy beam$^{-1}$ at a angular resolution of 0.32$''$ $\times$ 0.27$''$
 with a PA of $-$65$^\circ$ for both images. The ALMA theoretical rms noise for this configuration, integration time, and frequency is about 43 $\mu$Jy beam$^{-1}$,
 which is very close to the value we obtain in the continuum images.  For the line image rms noise we obtained a value of 6.5  mJy beam$^{-1}$ km s$^{-1}$ at a angular 
 resolution of 0.31$''$ $\times$ 0.26$''$ with a PA of $-$64$^\circ$. The ROBUST parameter was set also to natural.
The ALMA theoretical rms noise for this configuration, integration time, and frequency is about 4.3 mJy beam$^{-1}$,
 which is very close to the value we obtain in the line images. 
 Phase self-calibration was attempted on the continuum images but the resulting maps did not improved significantly, 
 so at this point we did not used the self-calibrated maps.
 We did not include the channel maps because they do not add extra information from that obtained from the moments maps.

\section{Results and discussion}

\subsection{HH 901}

In Figure \ref{f1} we present the {\it HST/ACS}  image from \citet{smi2010} of the 
optical jet HH 901 overlaid with the ALMA high velocity CO(2$-$1) and the 1.3 mm continuum emission. The CO emission reveals 
a bipolar and collimated outflow that is tracing the most inner part of the optical jet in an east-west orientation (with an approximate size of 5000 au). This map shows 
that the blue-shifted emission is located toward an eastern orientation, while red-shifted emission is to its western side. 
We integrate in radial velocities from $+$6.4 to $-$1.2 km s$^{-1}$ for the redshifted emission, and from $-$8.2 to $-$14.6 km s$^{-1}$ 
for the blueshifted emission for the CO outflow in the HH 901 object.  The CO emission between $-$1.2 km s$^{-1}$ to $-$8.2 km s$^{-1}$ 
is associated with the dusty pillar as is shown in Figure \ref{f2}. The LSR systemic velocity of the entire pillar associated with the object HH 901 
is about $-$5.0 km s$^{-1}$ (Figure \ref{f3}). The radial velocities of the innermost CO outflow are in agreement to those presented in \citet{rei2014}. 
They found, using slit positions of the optical [SII] and H$\alpha$ toward the HH 901, red-shifted emission in the west side of the flow 
with radial velocities about $+$25.0 km s$^{-1}$.  

\begin{figure*}
\centering
\includegraphics[scale=0.48]{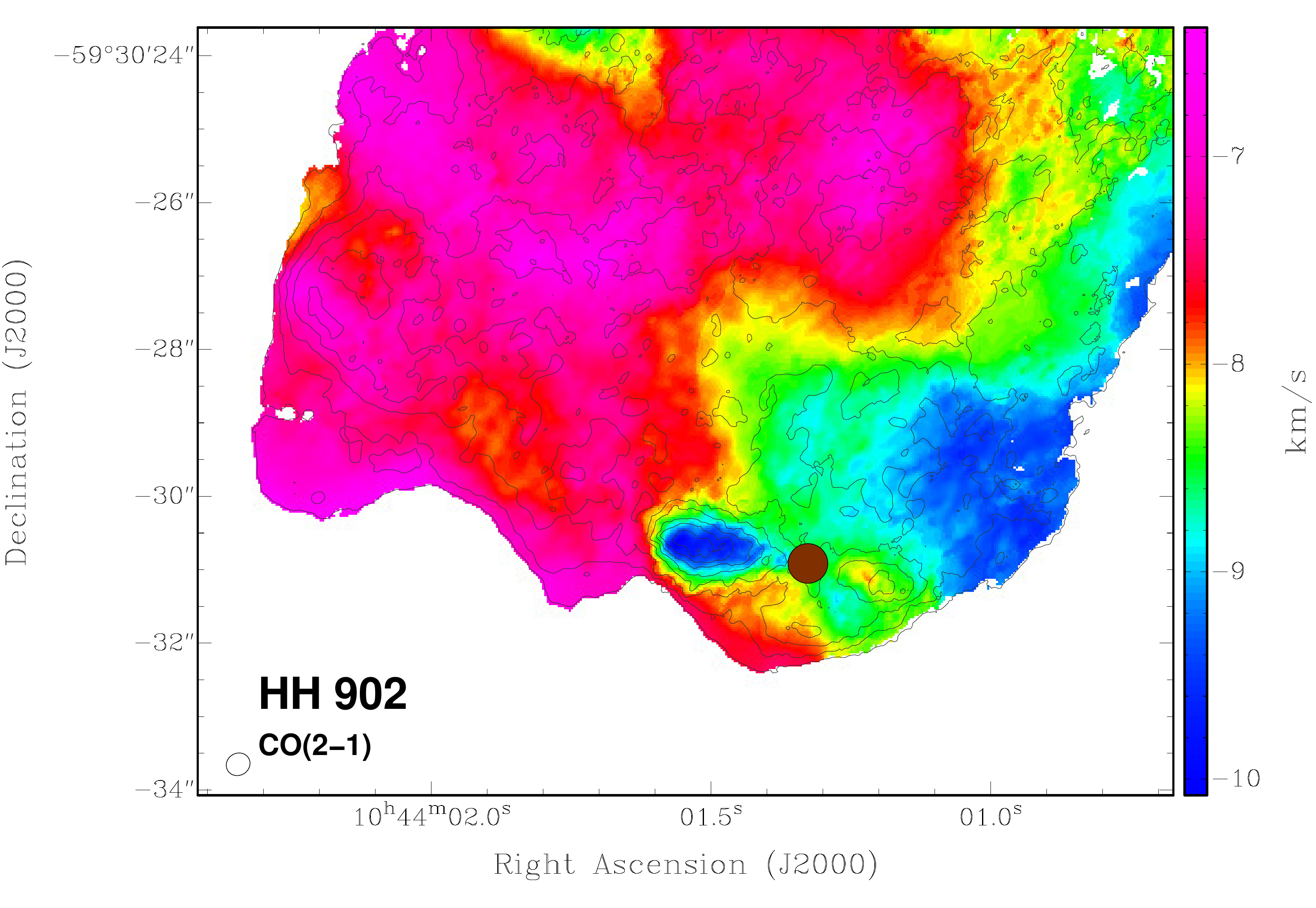}
\caption{ALMA CO(2$-$1) moment zero (contours) and one (colors) maps of the pillar and outflow HH 902. 
The contours range from 10\% to 90\% of the peak emission, in steps of 10\%. The peak of the millimeter 
line CO emission is 1.6 Jy Beam$^{-1}$ km s$^{-1}$. The half-power contour of the synthesized beam of the line image 
is shown in the bottom-left corner. The LSR radial velocity scale-bar is shown at the right. The brown circle
marks the position of the disk revealed in these ALMA observations for the HH 902 object.
For the CO moment zero image, we integrate in radial velocities from $-$15.2 to $+$3.7 km s$^{-1}$.
The LSR systemic velocity of the entire pillar associated with the object HH 902 is about $-$8.5 km s$^{-1}$.}
\label{f5}
\end{figure*}

 The 1.3 mm continuum image traces the disk and possibly part of the envelope at slightly larger scales. 
From a gaussian fitting to the continuum source from a map obtained using ROBUST parameter equal to 0.5 (yielding a better angular resolution), we obtained
 a deconvolved size of 0.49$''$ $\pm$ 0.03$''$ $\times$  0.33$''$ $\pm$ 0.03$''$
with a PA of 100$^\circ$ $\pm$ 7$^\circ$ and an integrated flux of 6.8 $\pm$ 0.4 mJy Beam$^{-1}$.
The corresponding physical sizes of these deconvolved values are about 700~au, which suggests that we are also seeing part of the envelope
and disk.  Assuming that the dust emission is optically thin and isothermal, the dust mass (M$_\mathrm{d}$) is directly proportional to the flux density (S$_\nu$)
 integrated over the source, as:

\begin{equation}
M_d=\frac{D^2 S_\nu}{\kappa_\nu B_\nu(T_d)},
\label{eq1}
\end{equation}

\noindent
where $D$ is the distance to Carina Nebula  \citep[2.3$\pm$0.1 kpc,][]{sb2008}, $\kappa_\nu$ the dust
mass opacity, and B$_\nu(T_d)$ the Planck function for the dust
temperature T$_d$.  In reality, thermal dust emission is probably not optically thin, 
hence the estimated mass is considered to be the lower limit. 
Assuming a dust mass opacity ($\kappa_\nu$) of 0.015
cm$^2$ g$^{-1}$ (taking a dust-to-gas ratio of 100) appropriate for these wavelengths (1.3 mm) \citep{Oss1994}, 
a typical opacity power-law index $\beta$ = 1.2 (obtained from our ALMA data),
as well as a characteristic dust temperature of 50 K, we
estimated a lower limit for the mass of the 
most compact part ({\it i.e.}, not interferometrically filtered) of the disk and envelope system associated with HH 901
of about 0.1 M$_\odot$. The mass uncertainty could be very large (a factor of up to 3 or 4) given the uncertainty 
in the opacity and in the dust temperature.

We can also estimate the volumetric and column density following the procedure
described in \citet{her2014}.   We obtained a lower limit for the column density of 10$^{24}$ cm$^{-2}$ and a volumetric density of
10$^8$ cm$^{-3}$ for the dusty source associated with the HH 901 object. These values are observed in very dense regions 
as compact envelopes and circumstellar disks toward the nearest GMC, Orion \citep{tei2016,taka2013}.   

If we assume that the continuum emission at these millimeter wavelengths is mostly arising from the disk, 
with a very small contribution of the envelope, the 0.1 M$_\odot$ could be linked to the disk.
If we further adopt a value M$_\mathrm{star}$/M$_\mathrm{disk}$ between 1 and 10 \citep{bate2018}, we can estimate that 
the protostar in the middle of the HH 901 object should be a low-mass star.
Furthermore, if the envelope is even more massive and the disk mass is smaller, 
we will have a protostar of lower mass. 
However, given the uncertainties on the mass estimation for the disks (a factor of 3 or 4), it could also be possible that the star in the middle
is an intermediate-mass star. This is in agreement with recent observations \citep{rei2013,rei2014}.

In Figures \ref{f2} and \ref{f3} show the kinematics and distribution of the molecular gas using the C$^{18}$O,  N$_2$D$^{+}$, and
$^{12}$CO spectral lines. These Figures present the moment zero emission from $^{12}$CO together with the moment one
of the C$^{18}$O,  N$_2$D$^{+}$, and $^{12}$CO lines. In order to compute these moments we integrate in radial velocities
from $-$14.6 to $+$6.4 km s$^{-1}$, which includes the emission from the pillars and the outflows. The molecule that better 
traces the pillars and the outflows is the $^{12}$CO, see Figure \ref{f1}.  From this Figure, one can see how the pillar is very well 
defined and with a clear east-west velocity gradient going from the blueshifted to redshifted, respectively. In the southern most
side of the pillar, the bipolar outflow HH 901 is evident. The bipolar outflow is only observed in the area traced
by the pillar probably because the critical density for the CO(2$-$1) ($\sim$10$^4$ cm$^{-3}$) decreases drastically outside
of the pillar and the outflow emission becomes too faint at larger spatial scales to be detected.
Both lines, the C$^{18}$O and N$_2$D$^{+}$ confirmed the east-west velocity gradient 
clearly revealed by the CO, as seen in Figure \ref{f3}. However, for the case of N$_2$D$^{+}$, this molecule is only observed in a small solid angle  
coincident with the 1.3 mm continuum image, probably tracing the densest parts of the pillar where the star formation is taking place.
The critical density of N$_2$D$^+$ J=3$-$2 is about 3$\times$10$^6$ cm$^{-3}$, which explains the fact that the N$_2$D$^+$ line 
has better spatial correlation with the 1.3 mm dust continuum emission. \citet{mat2019} reported that the N$_2$D$^+$ line is strongly
depleted towards the dusty massive core MM5/OMC-3 and this is explained in terms of chemical evolution. Once CO evaporates in 
the gas phase, molecules like N$_2$H$^{+}$ or N$_2$D$^{+}$ will not have an efficient formation process. This is, however, not observed
in our maps probably because it is needed a better angular resolution or sensitivity. 
  
\begin{figure*}
\centering
\includegraphics[scale=0.49]{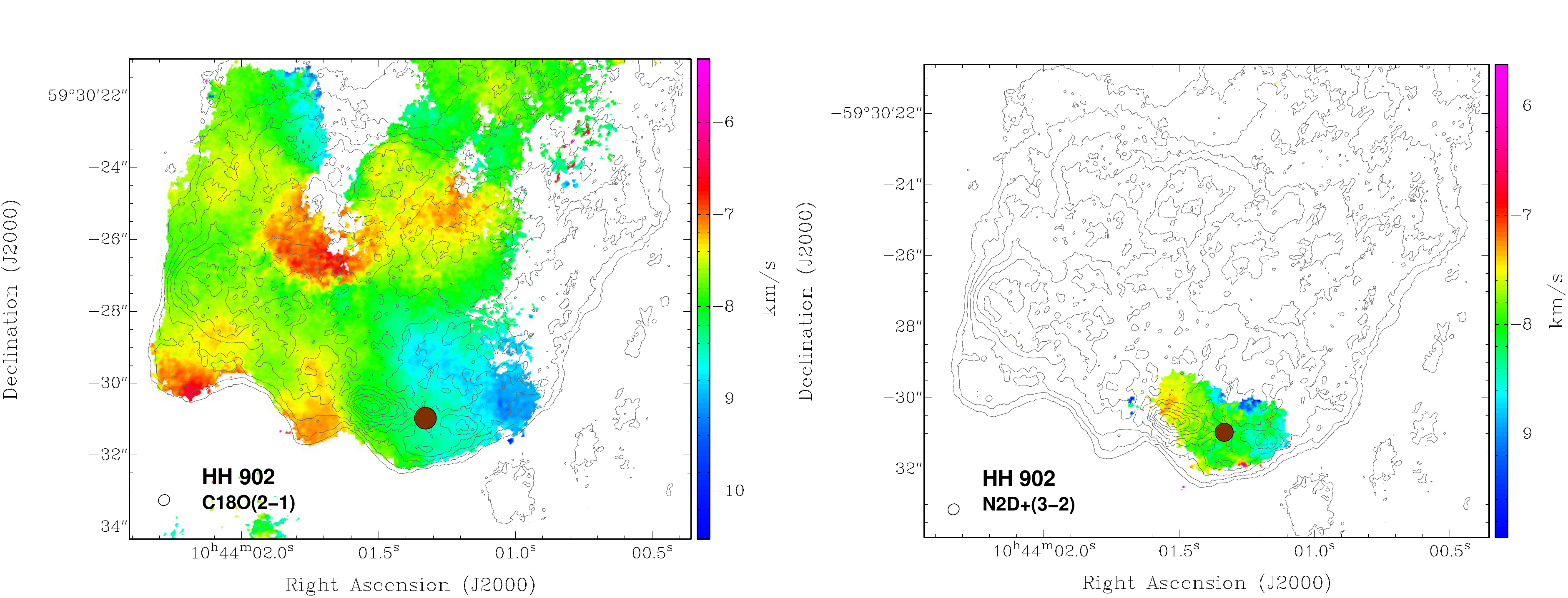}
\caption{ALMA C$^{18}$O(2$-$1) (left panel) and N$_2$D$^{+}$(3$-$2) (right panel) moment one maps overlaid 
with a contour image of the CO(2$-$1) moment zero of the pillar and outflow HH 902. 
The contours range from 10\% to 90\% of the peak emission, in steps of 10\%. The peak of the millimeter 
line CO emission is 1.8 Jy Beam$^{-1}$ km s$^{-1}$. The half-power contour of the synthesized beam of the line image 
is shown in the bottom-left corner of each panel. The LSR radial velocity scale-bar is shown at the right. The brown circle
marks the position of the disk$+$envelope revealed in these ALMA observations for the HH 902 object.
}
\label{f6}
\end{figure*}

As the CO(2$-$1) is probably optically thick, we can use the C$^{18}$O to estimate the opacity ($\tau_0$) and then to estimate 
the mass of the pillar and the outflow.  Assuming Local Thermodynamic Equilibrium, we can estimate the masses 
using the equations presented in \citet{sco1986,palau2007} for the J=2$-$1 transition, obtaining:

\tiny
\begin{equation}
\left[\frac{M_{H_2}}{M_\odot}\right]=1.2 \times10^{-15}\,T_\mathrm{ex}\,e^{\frac{16.59}{T_\mathrm{ex}}}
\,X_\frac{H_2}{CO} 
\left[\frac{\int \mathrm{I_\nu dv}}{\mathrm{Jy\,km\,s}^{-1}}\right]
\left[\frac{\theta_\mathrm{maj}\,\theta_\mathrm{min}}{\mathrm{arcsec}^2}\right]
\left[\frac{D}{\mathrm{pc}}\right]^2. 
\label{eq2}
\end{equation}
\normalsize

We used 2.8 as mean molecular weight,  an abundance ratio between the molecular hydrogen and the carbon 
monoxide ($X_\frac{H_2}{CO}$) of $\sim$10$^4$ \citep{sco1986}, an excitation temperature ($T_\mathrm{ex}$) of 50~K, 
an average intensity ($I_\nu$) of 0.01 Jy Beam$^{-1}$ for the outflow and 0.06 Jy Beam$^{-1}$ for the pillar,  
we take a velocity range ($dv$) of 5.5 km s$^{-1}$ for the outflow and 2.5 km s$^{-1}$ for the pillar, 
and a distance ($D$) to the source of 2.3$\pm$0.1~kpc. We then estimate a gas mass for the outflow of $\sim$ 5 $\times$10$^{-4}$ 
M$_\odot$ and for the pillar of $\sim$ 0.1 M$_\odot$.  For the mass of the outflow, we can correct this value for 
the inclination, therefore the true velocity range is $dv_{real}$ = $\frac{dv}{\sin \theta}$, where $\theta$ is the inclination angle taken to be
20$^\circ$ \citep{rei2013,rei2014}, so the $dv_{real}$ = 16 km s$^{-1}$ and the mass is 10$^{-3}$ M$_\odot$.

The uncertainty in the values of the mass arises mainly from the error in
the distance, which is only 4\%, the flux density which is about 10\% and the excitation temeperature. 
Given that the C$^{18}$O intensity line is approximately five 
times weaker than the $^{12}$CO line, we estimate an average opacity ($\tau_0$) of 0.2, 
so we corrected the mass value by a factor of $\left[\frac{\tau_0}{1-\mathrm{exp}(-\tau_0)}\right]$.
Furthermore, this mass estimation is in agreement with the mass obtained for the HH 901 pillar 
using a volumetric density of 10$^5$ cm$^{-3}$ and a physical volume of 10$''$ $\times$ 2$''$ $\times$  2$''$ obtained
in \citet{rei2013} and with a value of about 0.1 M$_\odot$.   

 For the outflow, the resulting kinetic energy (E$_k$ =$\frac{1}{2} m v^2$) is about $10^{42}$ ergs and the momentum ($p$=$mv$) is  
 1 $\times$ $10^{-2}$ M$_\odot$ km s$^{-1}$. 
 These values are slightly higher than those estimated by \citet{lum2014,zap2015,zap2018} for outflows associated with 
 low-mass protostars, but possibly consistent if the correction for outflow inclination is taken into account in these studies. 
 If we estimate the mechanical force for the outflow 
 F$_\mathrm{CO}$, we obtain a value of 1.5 $\times$ $10^{-5}$ M$_\odot$ yr$^{-1}$ km s$^{-1}$. This value is low compared from those 
 obtained for intermediate-mass young stars reported in Table 7 of \citet{van2016}, again suggesting that this outflow is associated
 with a low-mass or in the limit of an intermediate star.  If we add the momentum and energy from the optical counterpart 
 could be an intermediate-mass star.  Here, it is important to mention that the outflows in Carina are strongly irradiated,
  and maybe most of the energy budget is already in the optical irradiated counterparts, thus that a direct comparison with
  van Kempen's outflows could be not so straight.

As already mentioned in Section 1, \citet{rei2013,rei2014} suggested that the HH 901 jet is energized 
by a relatively young intermediate-mass star (2-8 M$_\odot$). However, this estimation of the mass (and evolutionary stage)
is very poorly constrained because they have not direct detection of the IR flux from the driving source of HH 901 \citep{rei2014}. 
This assumption is only inferred by the large mass-loss rates of a few 10$^{-6}$ M$_\odot$ yr$^{-1}$ (these are about of 
two orders of magnitude larger than those observed in T-Tauri stars) obtained from the optical ionized lines. 
These large mass-loss rates can be later transformed to mass-accretion rates and then to luminosity of the exciting star. 
However, the possibility that the powering source of the HH 901 is of intermediate-mass is in agreement 
with our ALMA data, but this intermediate-mass star must be on the low range of mass as suggested by the 
outflow energy. 

Given the high the mass-lost rate reported in the HH 901 object (10$^{-6}$ M$_\odot$ yr$^{-1}$), 
the protostar in the middle could be Class 0/I object, so this should be very obscured even in the IR wavelengths.
This is in agreement with the fact that the protostar is still surrounded by an envelope traced by the 1.3 mm 
and N$_2$D$^{+}$ emission. 
 
Following \citet{del2016}, we can estimate the photo-evaphoration time-scale for the pillar or globulette associated 
with the HH 901 object.  Taken the Equation \ref{eq2} (already discussed in that paper,) and which allow us to estimate the 
theoretical value of the globule/disk mass photo-evaporation rate \({\dot M}\) by Extreme Ultraviolet radiation (EUV): 

\tiny
\begin{equation}
  \dot{M}\EUV \simeq 
    4\times 10^{-9} \ 
  \left( \frac{F\EUV}{10^{10} \mathrm{\ s^{-1}\ cm^{2}}}  \right)^{1/2} \
  \left( \frac{R}{100 \mathrm{\ AU} } \right)^{3/2} 
  \ M_\odot \mathrm{\ year^{-1}},
  \label{eq3}
\end{equation}
\normalsize

where F$_\mathrm{EUV}$ is EUV flux, and R is the radius of the globule or disk. As the massive cluster Tr 14 is the closest 
(1.0 pc) to the HH 901 object, we ignored at this point the radiation for the massive cluster located also in this region Tr 16.  
The Tr 14 cluster has an UV luminosity \((Q_H)\) of \(2 \times 10^{50} \mathrm{\ photon\ s^{-1}}\) \citep{sb2008}, 
implying \(F\EUV= 10^{12} \mathrm{\ photon\ s^{-1}\ cm^{-2}}\) at the distance of the HH 901 object. For a radius of 1000 au, 
and using the Equation \ref{eq3}, we obtain a mass photo-evaporation rate \({\dot M}\) equal to 1 $\times$ 10$^{-6}$  M$_\odot$  year$^{-1}$.
Therefore, taking a mass of 0.1 M$_\odot$ for the pillar HH 901, we obtain that the pillar will be photo-evaporated 
in around 10$^5$ years. This timescale is shorter to that estimated, for example, for the massive EGG or globulette 
105-600 \citep{del2016} of about 10$^6$ years. However, the 105-600 globulette is farther away from the Tr 14 massive 
cluster (17 pc), and is more massive (0.1 to 0.3 M$_\odot$), which explains the large difference.  
 
 \subsection{HH 902}

In Figure \ref{f4} we present the {\it Hubble Space Telescope}/Advanced Camera for Surveys image from \citet{smi2010} of the 
optical jet HH 902 overlaid with the ALMA high velocity CO(2$-$1) and the 1.3 mm continuum emission. As in the HH 901 object (Figure \ref{f1}),
the CO emission also reveals a bipolar and collimated outflow that is tracing the most inner part of the optical jet in a north-west and south-east 
orientation. This map reveals that the blue-shifted emission is located toward an eastern orientation, while red-shifted emission is to its western side,
similar to the HH 901 object. We integrate in radial velocities from $+$3.7 to $-$5.6 km s$^{-1}$ for the redshifted emission, and from $-$10.8 to $-$15.2 km s$^{-1}$ 
for the blueshifted emission for the CO outflow in the HH 902 object.  The CO emission between $-$5.5 km s$^{-1}$ to $-$10.7 km s$^{-1}$ 
is associated with the dusty pillar as is shown in Figure \ref{f2}. The LSR systemic velocity of the entire pillar associated with the object HH 902 
is about $-$8.5 km s$^{-1}$. The radial velocities of the innermost CO outflow are in agreement to those presented in \citet{rei2014}. 
 
The 1.3 mm continuum emission in Figure \ref{f4} is also tracing, as in Figure \ref{f1}, the disk+envelope associated with the HH 902 object.  
However, in this Figure two objects, named A and B, are revealed. 
The component A is associated with HH 902 object, while the object B is to its north-west.
The source B is not associated with any molecular outflow, maybe because it is still in its pre-stellar phase. This source is extended with a 
size of about 2000 au and has a flux density of 38 $\pm$ 2.0 mJy Beam$^{-1}$. Assuming a dust temperature of 30 K, this flux density 
corresponds to a mass of 1.5 M$_\odot$, which is likely associated with an extended envelope.  For the millimeter component A, 
from a gaussian fitting to the continuum source from a map obtained using ROBUST equals to 0.5 (with a better angular resolution) 
we obtained a deconvolved size of 0.34$''$ $\pm$ 0.02$''$ $\times$  0.28$''$ $\pm$ 0.02$''$
with a PA of 107$^\circ$ $\pm$ 13$^\circ$ and an integrated flux of 6.5 $\pm$ 0.7 mJy Beam$^{-1}$. The corresponding physical sizes of these 
deconvolved values are about 700~au, which again suggests that we are also seeing part of the envelope and disk. 

Following the Equation \ref{eq1}  and taking similar values to the ones assumed for the object HH 901, we estimated a lower limit 
(because again the thermal dust emission is probably not optically thin) for the mass of the 
most compact part of the disk and envelope system (of source A) associated with the HH 902
of about 0.1 M$_\odot$. In a similar way as for the HH 901 continuum object, this mass for the dust  for the HH 902 object 
corresponds to a low-mass protostar.

In Figures \ref{f5} and \ref{f6}, the kinematics and distribution of the molecular gas using the C$^{18}$O,  N$_2$D$^{+}$, and
$^{12}$CO spectral lines are presented. These Figures present the moment zero emission from $^{12}$CO together with the moment one
of the C$^{18}$O,  N$_2$D$^{+}$, and $^{12}$CO lines. Similar to Figures \ref{f3} and \ref{f4}, in order to compute these moments we integrate in radial velocities
from $-$15.2 to $+$3.7 km s$^{-1}$, which includes the emission from the pillars and the outflow. The molecule that better 
traces the pillars and the outflow is again the $^{12}$CO, as seen in Figure \ref{f5}, but now for the HH 902 object.  From this Figure, one can see how the pillar is very well 
defined and with a clear east-west velocity gradient going from the blueshifted to redshifted, respectively. In the most southern
side of the pillar, the bipolar outflow HH 902 is evident.  Both lines, the C$^{18}$O and N$_2$D$^{+}$ confirm the east-west velocity gradient 
clearly revealed by the CO, as seen in Figure \ref{f5}. However, for the case of N$_2$D$^{+}$, this molecule again is only observed in small solid angle  
coincident with the 1.3 mm continuum image, probably tracing the densest parts of the pillar, as in the HH 901 object.

Using the Equation \ref{eq2} and assuming similar values for the excitation temperature, an abundance ratio, and distance, we can obtain the mass for the outflow and the pillar
associated with the HH 902 object. We take an average intensity ($I_\nu$) of 0.015 Jy Beam$^{-1}$ for the outflow and 0.07 Jy Beam$^{-1}$ for the pillar,  
and we also take a velocity range ($dv$) of 5.0 km s$^{-1}$ for the outflow and 3.0 km s$^{-1}$ for the pillar. For these values,  we estimate a gas mass 
for the outflow of  7 $\times$ 10$^{-4}$ M$_\odot$ and for the pillar of $\sim$ 0.2 M$_\odot$. We also corrected these mass values by the opacity.
Moreover, correcting these values by the inclination, we obtain a mass for the outflow of 10$^{-3}$ M$_\odot$.
 
Taking again the Equation  \ref{eq3}  and similar values for the EUV radiation (\(F\EUV= 10^{12} \mathrm{\ photon\ s^{-1}\ cm^{-2}}\)) at the distance of the HH 902 object
and a radius of 4 $\times$ 10$^3$ au, we obtain a mass photo-evaporation rate \({\dot M}\) equal to 1 $\times$ 10$^{-5}$  M$_\odot$  year$^{-1}$.
Therefore, taking a mass of 0.2 M$_\odot$ for the pillar HH 902, we obtain that the pillar will be photo-evaporated 
in around 2 $\times$ 10$^4$ years, which is a lower value to the timescale of the HH 901.

Given the short photo-evaporated timescales of the molecular pillars and that the protostars in these pillars are still very embebed, 
we suggest that that the disks inside of the pillars will be quickly affected by the radiation of the massive stars (in about 10$^4$ -- 10$^5$ years), 
forming proplyds like those observed in Orion \citep{ode1993, ode2008, hen1999}.

 \section{Conclusions}
 
 In this study, we present sensitive and high angular resolution observations of the famous HH 901 and HH 902 objects located in the Carina Nebula using
ALMA.  The main results of our work can be summarized as follows.
 
 \begin{itemize}
 
 \item Our observations reveal for the first time, the bipolar CO collimated outflows, and the dusty disks (plus envelopes) that are energizing the extended and irradiated HH 901/902 
objects far from the pillars.

 \item We find that the masses of the disks$+$envelopes are about 0.1 M$_\odot$ and that of the bipolar outflows are between 10$^{-3}$ -- 10$^{-4}$ M$_\odot$, which suggests that
the outflows could be energized by low- or  maybe intermediate- mass protostars. 
 
 \item  We also reveal the kinematics of the gas in the pillars together with their respective gas masses (0.1 -- 0.2 M$_\odot$).
 
 \item We estimate that the pillars will be photo-evaporated in around of 10$^4$ to 10$^5$ years by the massive and luminous stars located in the Tr 14 cluster.

 \end{itemize}
 

\acknowledgments
We are very thankful for the thoughtful suggestions of the anonymous referee that helped to improve our manuscript.
This paper makes use of the following ALMA data: ADS/JAO.ALMA\#2017.1.00912.S. ALMA is a partnership of ESO (representing its member states), 
NSF (USA) and NINS (Japan), together with NRC (Canada), MOST and ASIAA (Taiwan), and KASI (Republic of Korea), in cooperation 
with the Republic of Chile. The Joint ALMA Observatory is operated by ESO, AUI/NRAO and NAOJ.
This research has made use of the SIMBAD database, operated at CDS, Strasbourg, France. L.A.Z. and J. A. T. are grateful to CONACyT, M\'exico, and DGAPA, 
UNAM for the financial support.


\vspace{5mm}
\facilities{HST(STIS), ALMA}

\software{CASA\citep{mac2007}, KARMA\citep{goo1996}}





\end{document}